# Dirac Cones and Room Temperature Polariton Lasing Evidenced in an Organic Honeycomb Lattice

*Simon Betzold\*, Johannes Düreth, Marco Dusel, Monika Emmerling, Antonina Bieganowska, Jürgen Ohmer, Utz Fischer, Sven Höfling, and Sebastian Klembt\**

S. Betzold, J. Düreth, M. Dusel, M. Emmerling, S. Höfling, S. Klembt

Julius-Maximilians-Universität Würzburg, Physikalisches Institut and Würzburg-Dresden Cluster of Excellence ct.qmat, Lehrstuhl für Technische Physik, Am Hubland, 97074 Würzburg, Germany

A. Bieganowska

Wroclaw University of Science and Technology, Faculty of Fundamental Problems of Technology, Department of Experimental Physics, Wyb. Wyspiańskiego 27, 50-370 Wroclaw, Poland

J. Ohmer, U. Fischer

Julius-Maximilians-Universität Würzburg, Department of Biochemistry, Am Hubland, 97074 Würzburg, Germany

\*E-mail: simon.betzold@uni-wuerzburg.de, sebastian.klembt@uni-wuerzburg.de

**Artificial one- and two-dimensional lattices have emerged as a powerful platform for the emulation of lattice Hamiltonians, the fundamental study of collective many-body effects, and phenomena arising from non-trivial topology. Exciton-polaritons, bosonic part-light and part-matter quasiparticles, combine pronounced nonlinearities with the possibility of on-chip implementation. In this context, organic semiconductors embedded in microcavities have proven to be versatile candidates to study nonlinear many-body physics and bosonic condensation, and in contrast to most inorganic systems, they allow the use at ambient conditions since they host ultra-stable Frenkel excitons. We implement a well-controlled, high-quality optical lattice that accommodates light-matter quasiparticles. The realized polariton graphene presents with excellent cavity quality factors, showing distinct signatures of Dirac cone and flatband dispersions as well as**

**polariton lasing at room temperature. This is realized by filling coupled dielectric microcavities with the fluorescent protein mCherry. We demonstrate the emergence of a coherent polariton condensate at ambient conditions, taking advantage of coupling conditions as precise and controllable as in state-of-the-art inorganic semiconductor-based systems, without the limitations of e.g. lattice matching in epitaxial growth. This progress allows straightforward extension to more complex systems, such as the study of topological phenomena in two-dimensional lattices including topological lasers and non-Hermitian optics.**

1. Introduction

Increasing technological control over the last decade has allowed increasingly sophisticated implementations of artificial lattice systems for the study and exploration of novel and exciting states of matter[1,2] as well as the experimental emulation of complex lattice Hamiltonians.[3] Inspired by exciting results in the field of ultracold atoms[4,5] and ion traps,[6] coupled optical lattices have emerged as a comparatively simple, on-chip platform for lattice emulation[7-9]. In this context, coupled high-quality vertical resonator microcavities are widely used as a precise and well-controlled optical platform for lattice emulation and the realization of complex lattice Hamiltonians.[10-12] By placing a suitable emitter material between two highly reflective mirrors, the emission and reabsorption of light can become dominant, resulting in a strong coupling of light and matter. Because of this, new eigenstates, called exciton-polaritons (polaritons) emerge.[13,14] Interestingly, they inherit large optical nonlinearities from their matter part, while retaining a low effective mass and great experimental accessibility form their light part. Landmark results in this system include Bose-Einstein condensation,[15,16] superfluidity[17] or electrically pumped polariton laser devices.[18] Placing such *quantum fluids of light*[19] in an optical lattice potential landscape using a variety of techniques[11] has led to the realization of lasing from a topological defect,[20,21] the implementation of a spin *XY*-Hamiltonian,[12] and the fabrication of polariton graphene[22-25] including the realization of a polariton topological Chern insulator.[26] The main material platform has traditionally been high-quality Gallium-Arsenide (GaAs)-based crystalline semiconductor microcavities, which benefit from a very mature epitaxial growth and nanofabrication.[11] While the platform even allows for electrical operation of polariton devices,[18,25,27,28] the fabrication requires state-of-the-art clean room facilities and operation is typically limited to cryogenic temperatures.

Novel emitter materials such as organic emitters,[29-32] transition metal dichalcogenides[33-36] and perovskites[37-40] have been successfully introduced into optical (dielectric) microcavities, adding a comparatively simple fabrication process as well as room temperature stable excitons to the equation. While respectable attempts have been made to realize two-dimensional polariton lattices hosting room-temperature emitter materials[41-43], the combination of a versatile platform with high quality factor (Q-factor), low disorder, and tunable and controllable inter-site couplings remains challenging.

In this work, we present a high-quality two-dimensional polariton lattice that hosts a room-temperature stable organic emitter material with excitons. A graphene-like honeycomb lattice is realized by nanofabricating dimples into a borosilicate substrate by focused ion beam milling. The dimpled lattice is subsequently oversputtered with pairs of $SiO_2$ and $TiO_2$ layers to form a dielectric, highly reflective distributed Bragg reflector (DBR). Finally, the fluorescent protein mCherry, closely related to the famous green fluorescent protein,[44] is drop-cast onto an identical, planar DBR and mechanically sandwiched between the patterned honeycomb mirror and the planar mirror. The resulting structure is a high-quality microcavity lattice that confines light between the two mirrors (z-direction). In addition, the light is confined in-plane (x,y direction) by the hemispherical dimple cavities.[45] The distance and curvature of the dimples are designed so that the inter-site coupling is large and the lattice is dominated by the nearest-neighbor coupling. We observe room temperature exciton polaritons in a very well-defined graphene-like band structure with Dirac cone dispersions and flatbands, as well as clear observation of polariton lasing.

## 2. Results

A detailed schematic representation of the fabricated device is shown in **Figure 1a**. It consists of two DBRs, each sputtered onto borosilicate glass substrates with a roughness of less than 1 nm RMS, between which a thin film of the fluorescent protein mCherry is sandwiched. Prior to coating the upper substrate with the mirror pairs, focused ion beam milling (FEI Helios NanoLab DualBeam) was used to introduce arrays of hemispheric dimples, arranged in the distinct geometry of a honeycomb lattice. The individual sites have diameters between 3 μm and 5 μm and depths between 100 nm and 380 nm, depending on the array. These hemispheres ensure that the light field of the fundamental mode is spatially confined with an effective radius of about 1.0-1.5 μm. Subsequently, 9 mirror pairs of $SiO_2/TiO_2$ were deposited on the substrate by ion beam sputtering (Nordiko 3000). The individual layer thicknesses were chosen so that the stopband center is at a wavelength of $\lambda_C \approx 630$nm and the first, high-energy Bragg minimum is at 532nm, allowing for efficient laser excitation. For an atomic force microscopy (AFM) measurement of this structured mirror, see **Figure S1**.

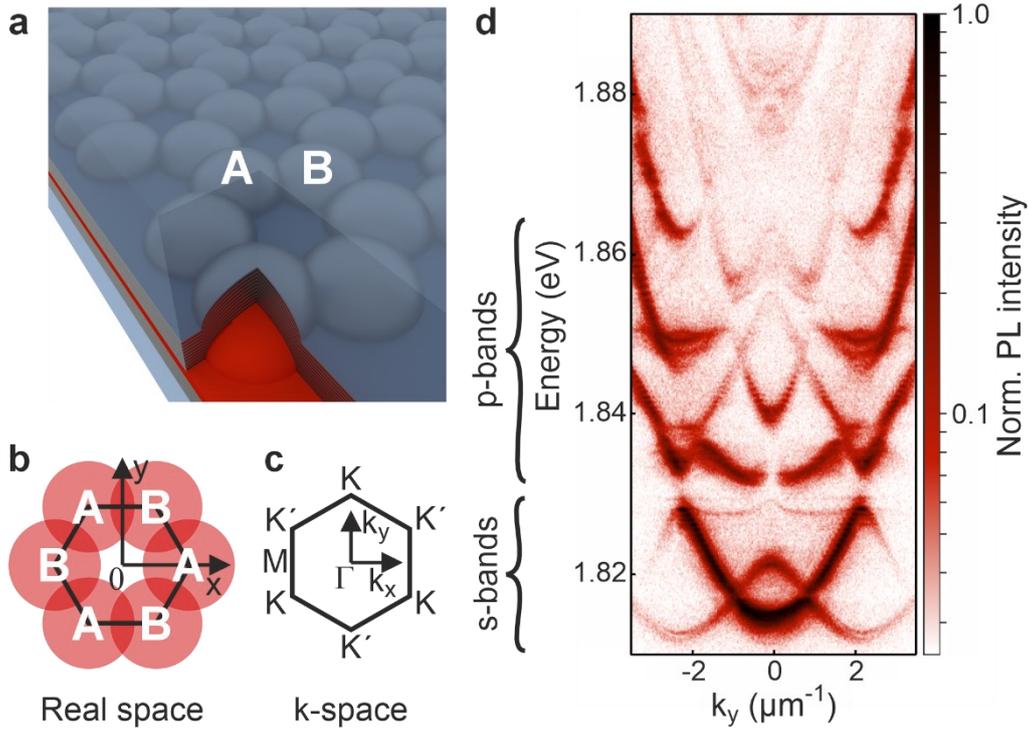

**Figure 1.** a) Schematic representation of the room-temperature polariton graphene, consisting of the fluorescent protein mCherry. A and B denote the two-site unit cell of the lattice. For the sample preparation, the highly concentrated solution was coated onto a planar DBR and then mechanically sandwiched between the planar and the structured DBR. b) Schematic representation of the real-space lattice as well as c) the geometry in k-space showing the K- and K'-points of the lattice. d) Photoluminescence dispersion measurement in K-K' direction.

A highly concentrated solution (200 g/l) of mCherry was coated onto an identical, planar DBR. The patterned DBR was then flipped and pressed onto the mCherry, mechanically introducing the fluorescent proteins into the dimples. Finally, the device is dried for 48 hours under a constant pressure of about 0.25 N/cm$^2$. See **Section 4** and **Figure S2** for details on sample preparation. The structure yields an experimental microcavity Q-factor of 4830 for individual traps, a Q-factor of about 1200-1400 for the two-dimensional system, and a cavity with an optical thickness of 1100 nm and a Rabi splitting of 318 meV in the planar area next to the lattice, which is in line with previous publications.[45,46] For more details, see **Figure S3**. To ensure the approximate length of the cavity as well as the parallelism of the two mirrors, we use built-in spacers. The real-space and k-space geometric properties of the honeycomb lattice are shown in **Figure 1b** and **1c,** respectively. The K- and K'-points are clearly related to the occurrence of Dirac cone dispersions in this lattice,[47] which is visible in the angle-resolved photoluminescence (PL) measurement displayed in **Figure 1d**. The measurement was performed using a 532 nm continuous wave diode laser which was focused to a Gaussian spot with a size of approximately 20μm in the center of the lattice. For all measurements shown, a lattice of hemispheres with a diameter of 4 μm and a center-to-center distance of 2 μm was employed. In the lowest S-band at about ±1.0 μm$^{-1}$, the dispersion is well described by a linear crossing of the modes. Above the S-band ranging from 1.81-1.83 eV, the P-band and higher order bands are visible. Even small details of the band structure can be well resolved due to the excellent combination of quality factor and coupling in the lattice.[23, Fig. 1d]

To study the organic room-temperature polariton graphene in more detail, we use a photoluminescence tomographic technique to access the full band structure in the in-plane momentum $k_{x,y}$ and in energy (PL emission wavelength). **Figure 2a-d** show PL dispersion measurements in different high-symmetry directions referred to in Figure 1c. See **Section 4** for more details on the measurement methods. Some features of the S- and P-bands of the measured dispersion relation, like the missing anti-binding S-band in Figure 2a at $k_x$=0 μm$^{-1}$ along K-Γ-K', are governed by sublattice interference. This band, however, is well visible in the second Brillouin zone shown in Figure 2b.[23] Figure 2c displays the dispersion along the K-K direction, highlighting the iconic Dirac cone, while the latter is expectedly absent in Figure 2d when measuring along the M-Γ-M direction. Due to the fully available k-space dispersion information (within the light cone), we can reconstruct and plot a reduced Brillouin zone in **Figure 2e**. We show the mode energies along the high symmetry points Γ, K/K' and M. The Dirac cone at K and the open bands at M are clearly visible and well described by a tight-binding model considering nearest- and next-nearest-neighbor coupling (red line). In addition, the flatband of the P-mode is visible above the S-band. It is formed by destructive interference of the overlapping local modes in real-space and appears with a vanishing band curvature. **Figure 2f** displays the Dirac cones at the K/K'-points at the intersection energy, similar to well-known angle-resolved photoemission spectroscopy measurements of graphene [48] (with μm$^{-1}$ instead of Å$^{-1}$), while **Figure 2g** shows the Γ-points of the 2$^{nd}$ Brillouin zone. Similar to the k-space tomographies, we map the real-space potential landscape by incrementally shifting the image with respect to the entrance slit of our spectrometer, accessing the PL emission energy as a function of (x,y). To make a qualified statement about the homogeneity of the potential landscape, we plot the energy of the anti-binding S-band $S_{AB}$ in **Figure 2h**. For this purpose, the energy range between 1.818eV and 1.830eV was spectrally summed. As expected for the anti-binding mode, the luminescence is located at the individual sites, highlighting the honeycomb pattern. The intensity is overall homogenous, following the Gaussian laser spot excitation with a diameter of about 20μm. In contrast, the P-flatband mode $P_{flat}$, which is energetically in the range of 1.830eV and 1.837eV, is located at the contact points of the lattice (as seen in **Figure 2i**) due to the dumbbell shape of the P-mode and the hybridization in the flatband.[23,49-51] Apart from the emission wavelength of about 685nm, the presented dispersions compare very well with the results obtained at cryogenic temperatures of typically T= 4K in single-crystal III/V-semiconductor-based structures.[23] However, while in coupled micro-resonator pillars the physical overlap of the sites creates the coupling, in our lattice the in-plane confinement is caused by the overlapping hemispheres in the top mirror. Looking more closely at the results in Figure 2, we see first that the S-band is energetically asymmetric around the Dirac energy of approximately $E_D$~1.817 eV, extending more to higher than to lower energies (cf. Figure 2a-c).

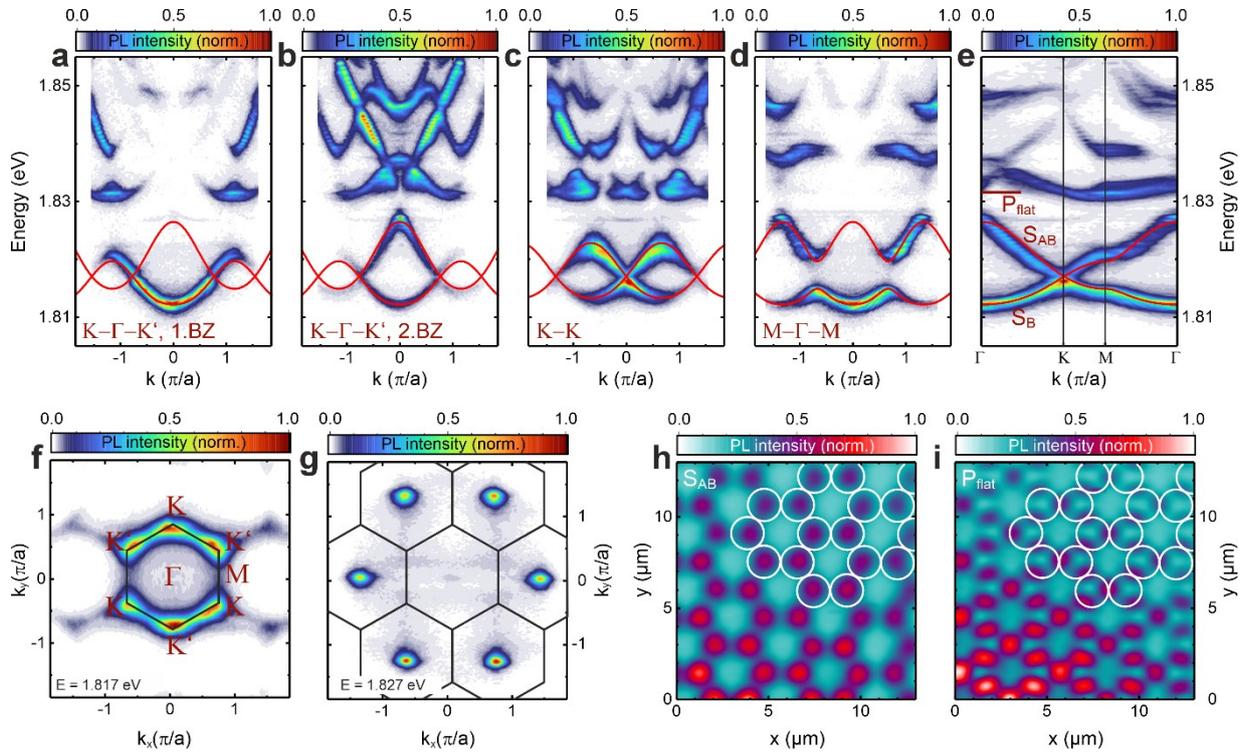

**Figure 2.** a)-d) Polariton dispersions along the K-Γ-K' (first and second Brillon zone), K-K and M-Γ-M directions of the Brillon zone. e) Representation of the reduced zone scheme calculated by averaging the first as well as an adjacent second Brillon zone. f)-g) Iso-energy cuts of the Fourier space at the energy of the Dirac cones at 1.817 eV (f) and at the energy of the top of the π*-band at 1.827 eV (g), which shows the occupation of the Γ-points in the second Brillon zone. h)-i) Real-space images of the anti-binding S subband (h), for which the intensity was spectrally summed between 1.818eV and 1.830eV, and of the lowest P subband (i), for which the intensity was spectrally summed between 1.830 and 1.837eV.

Second, some modulation of the flatband in Figure 2e can be observed. Both signatures arise from next-nearest-neighbor coupling, which is common in close-coupled systems. The energy of a bight-binding model of a Honeycomb lattice is given by

$$E(k) = \pm t_1 \cdot \sqrt{3 + f(k)} - t_2 \cdot f(k) \qquad (1)$$

where $k$ is the in-plane wave vector and $f(k)$ is the geometric function

$$f(k) = 2 \cdot \cos(\sqrt{3}k_y a) + 4\cos\left(\frac{\sqrt{3}}{2}k_y a\right)\cos\left(\frac{3}{2}k_x a\right). \qquad (2)$$

The coefficients $t_1$ and $t_2$ are associated with the relative strength of the nearest-neighbor coupling $t_1$ to the next-nearest-neighbor coupling $t_2$ and $a$ is the lattice constant. Fitting the dispersions, we get $t_2/t_1$=0.25meV/2.38meV=11%. It is now worthwhile to compare this result with existing ones in more conventional inorganic semiconductor-based polariton lattices. **Table 1** shows the tight-binding coupling data for several polariton lattices where all compared works have honeycomb geometries. It turns out that, although the resonator designs and the absolute coupling strengths are different, the ratio between next-nearest-neighbor coupling $t_2$ and nearest-neighbor coupling $t_1$ is very comparable in all configurations and material systems. This shows that our hemispherical microcavities are also competitive with established III-V based microcavities in this respect.

**Table 1.** Comparison of the tight binding coupling parameters of the present work with those existing in conventional inorganic semiconductor cavities.

|  | $t_1$ (meV) | $t_2$ (meV) | $t_2/t_1$ | Confinement technique |
|---|---|---|---|---|
| This work | 2.38 | 0.25 | 0.11 | Hemispheric top mirror |
| PhD Thesis Tristan H. Harder[a] | 0.356 | 0.044 | 0.12 | Etch-and-overgrowth technique |
| Jacqmin et al.[23] | 0.25 | 0.02 | 0.08 | Conventional pillar etching |
| Real et al.[52] | 0.18 | 0.014 | 0.08 | Conventional pillar etching |
| Whittaker et al.[24] | 0.12 | 0.008 | 0.07 | Conventional pillar etching |

[a]publication in preparation, see [53] for the PhD thesis.

This characteristic, together with the remarkable stability of Frenkel excitons in fluorescent proteins, sets the stage for exploring the nonlinear regime of polariton lasing in 2D lattices at room temperature, as demonstrated in our subsequent experiment. We excite the Honeycomb lattice again with a Gaussian spot of about 20 μm diameter. However, a wavelength-tunable optical parametric oscillator system using nanosecond pulses is now employed. This system is specifically tuned to 532 nm to resonate with the first Bragg minimum of the top mirror in our sample. As we systematically increase the pump power from $P = 0.12$ μJ/pulse (**Figure 3a**) to $P = 0.27$ μJ/pulse (**Figure 3b**), and finally to $P = 1.89$ μJ/pulse (**Figure 3c**), a remarkable qualitative change in the lattice spectrum is observed. First, a significant nonlinear enhancement of the emission intensity is observed, concentrated in the spectral region of the d-bands in our lattice. A nonlinear increase of three orders of magnitude is observed in the integrated intensity shown in **Figure 3e** (black dots). Furthermore, Figure 3e shows a significant reduction of the linewidth at the polariton condensation threshold $P_{th} \approx 0.3$ μJ/pulse, indicating the build-up of phase coherence. Finally, **Figure 3d** illustrates a persistent blueshift of the mode above the threshold which results from phase-space filling effects[54,55] and is attributed to the exciton-polariton nature of the system, resulting in an overall shift of about 0.6 meV. Several factors influence the spectral position of the dominant lasing mode, which results in lasing in a d-band mode in this device. Primary factors are the exciton-photon detuning, the overlap between the external gain and the Bloch modes, and the size of the pump spot. If the detuning is such that the modes are in the spectral range from 2.0 eV to 1.9 eV and the system is excited with a small pump spot, it tends to start lasing with its lowest energy mode.[45,54] As the exciton-photon detuning increases, lasing shifts to the higher order modes. In addition to the cavity length, the energy of the lasing mode is also influenced by the position and size of the pump spot. Thus, the energy can be shifted by about 20 meV using this approach.[45] Please note that the dip in the input-output curve at about 0.5 μJ/pulse in Figure 3e is caused by mode competition due to excitation with an extended laser spot.

To further explore the spatial coherence of this mode, we investigate its correlation function $g^{(1)}$. Using a Michelson interferometer in mirror-prism configuration, we overlap the real-space luminescence from the device with its mirror image on a beam splitter and combine them on a high-resolution CCD camera. This measurement is performed well above the threshold with an excitation power of about $P \approx 7P_{th}$. **Figure 3f** shows the resulting interference image, where a distinct interference pattern can be observed with significant spatially extended fringes, characteristic of an extended coherent mode. The spatially resolved magnitude of the coherence measurement shown in **Figure 3g** confirms that the coherent part extends over several unit cells, exceeding 10µm. A mode tomography of this mode below the threshold can be found in **Figure S4**. This specific Bloch mode likely arises from the interference of the $3d_{xy}$ modes of individual hemispherical potentials, leading to the highest emission intensity at the center of each hexagon, while destructive interference occurs in the center of each hemisphere. A similar signature was observed in a square lattice in an inorganic microcavity by Kim et al.[56]

These results not only demonstrate the potential of polariton lasing in our system, but also underscore the spatially extended coherence of the emergent coherent mode. The observed effects, including nonlinear enhancement, linewidth reduction, and persistent blueshift, contribute to a comprehensive understanding of the intricate exciton-polariton dynamics in our lattice. This level of control and coherence in a room temperature environment opens new avenues for practical applications and further advances in the study of nonlinear phenomena in condensed matter physics.

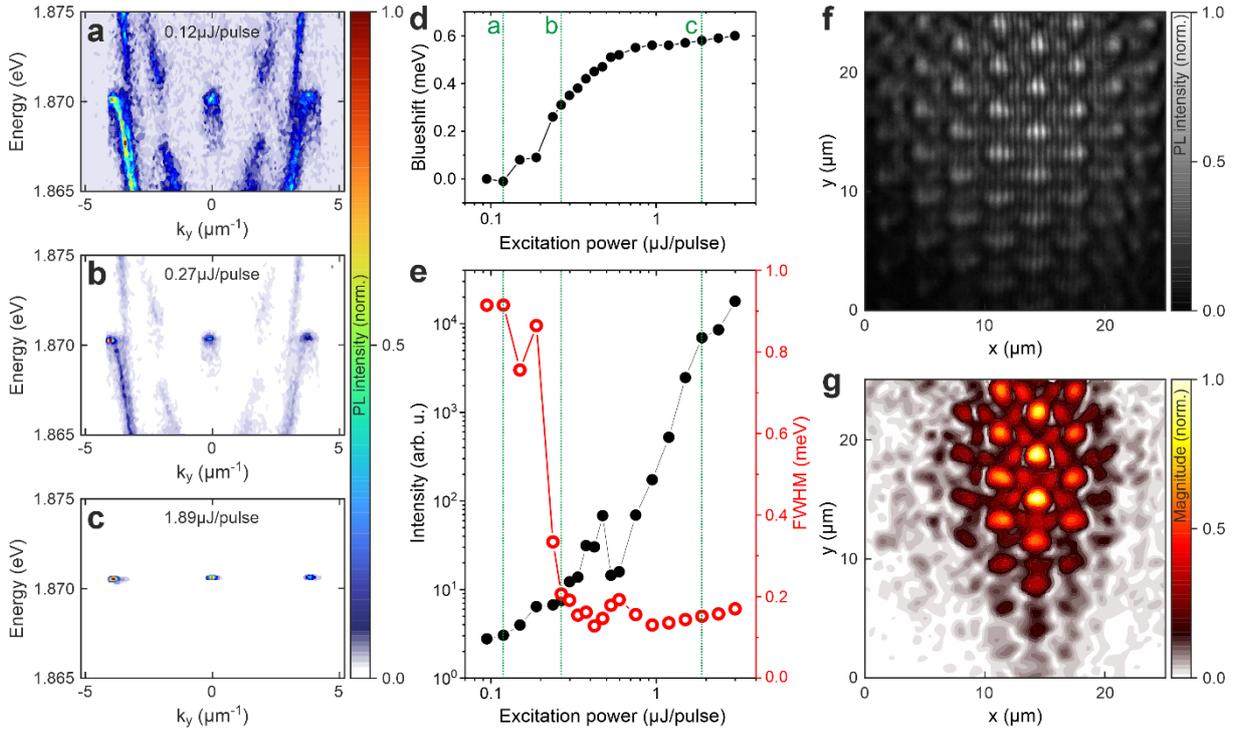

**Figure 3.** Polariton dispersions of the D-bands along the K-M-K' direction below the condensation threshold (a), at the threshold (b) and well above the condensation threshold (c) under pulsed excitation. D) shows the resulting energy shift for different excitation powers, while e) shows the corresponding input-output characteristics (black) and the linewidth (red). F) Michelson interference image of the polariton condensate at an excitation power of ~2 µJ/pulse and the corresponding spatial $g^{(1)}$-coherence map in (g), showing a spatial coherence well above 10µm and thus several unit cells.

## 3. Conclusion

In this work, we have successfully demonstrated a versatile platform to bring the physics of photonic lattices in general, and polaritonic lattices in particular, to the room temperature regime. We use an organic emitter material that hosts room temperature stable excitons to demonstrate polariton lasing in a complex potential landscape. The geometry of choice is the honeycomb lattice, as it has exciting connections to emulating graphene physics, topological insulators and lasers, flatband physics, and many more. The use of dielectric mirrors makes this approach completely independent of material constraints, as long as the emission wavelength and the mirror stopband match. While our experiments have been successfully performed on cavities where the emitter is sandwiched between the mirrors, this platform can easily be extended to open cavity approaches[57,58], enabling unprecedented freedom in the tuning and deterministic manipulation of research platforms and novel optoelectronic devices.

## 4. Experimental Section/Methods

*Sample preparation*: The glass substrates are cleaned by a general cleaning process before each step: Ultrasonic cleaning in water, acetone and isopropyl alcohol for 10 minutes each and subsequent oxygen plasma treatment for 2 minutes. After the general cleaning process, the top substrate is mounted in a FEI Helios Dual Beam system. Subsequently, the lattice is milled into the substrate with Gallium ions using an emission current of 6.5 nA, an acceleration voltage of 30 kV, a dwell time of 15 μs and 45 passes.

The spacers "mesas" on the bottom substrate are prepared by spin-coating photoresist onto the substrate and subsequent exposure with UV light, development and sputtering of 250 nm SiO2. The dielectric mirrors are deposited on the prepared bottom and top substrates by dual ion beam sputtering (Nordiko 3000), utilizing an assist ion source. In this case, 9 bilayers of $SiO_2/TiO_2$ with a thickness of 103.4 nm and 64.8 nm respectively are deposited. The last layer of $TiO_2$ is capped with 15 nm of $SiO_2$.

To assemble the cavity, both the bottom and top are cleaned. Then, 3.5 μl of mCherry solution in water (200 g/l) is pipetted onto the planar mirror and slightly spread around. The second, structured mirror is placed on the mCherry, pressed in place and left to dry under a force of about 0.25 N/cm$^2$ for 2 days. After this time, the substrate side of the top mirror is cleaned again to get rid of any contamination of mCherry on the top surface.

*Experimental setup*: In our experimental setup, we used two lasers for sample excitation. The first laser, a continuous wave diode laser emitting at 532 nm, was used for pre-characterization and band structure determination. For experiments in the nonlinear regime, we used a wavelength-tunable optical parametric oscillator system with nanosecond pulses, also tuned to 532 nm. The excitation beam was directed onto the sample surface using a high numerical aperture (NA = 0.42) objective (50x). The laser had a spot size of approximately 20 μm in diameter, which was achieved by focusing the laser on the back focal plane of the objective with a 30 cm focal length lens.

The emission from the sample was collected in reflection geometry through the same objective, filtered with a 550 nm long pass filter, and directed to the entrance slit of a 500 mm Czerny–Turner spectrometer. The signal was recorded with a Peltier-cooled EMCCD (Andor Newton 971). The spectrometer is equipped with three different gratings (150 lines/mm, 300 lines/mm, and 1200 lines/mm) and a motorized entrance slit, resulting in a spectral resolution of up to 200 μeV for energies around 2 eV. Most of the measurements were performed in a Fourier imaging configuration using an additional Fourier lens to collect angle dependent information in the back focal plane of the microscope.

A Michelson interferometer in a mirror-prism configuration was used to evaluate the coherence properties of the emission. The real-space image from the microcavity was split by a 50:50 beam splitter into two arms, one directed to a plane mirror and the other to a prism reflector reflecting the image back to the beam splitter. Interference between the images was recorded by a high-resolution imaging CCD camera (Andor Clara, 6.5 μm pixel size) at the output of the beam splitter.

For further information about the experimental setup, please see **Figure S5**.

*Mode tomography and hyperspectral imaging*: The imaging lens in front of the spectrometer is mounted on a motorized linear translation stage. This allows to take slices of the real or momentum space one after the other at a fixed x or $k_x$ value. Afterwards, they can be combined into a 4D matrix to a mode tomography (intensity for (x,y) vs. energy) or a hyperspectral image (intensity for ($k_x$,$k_y$) vs. energy). In this way, for example, an energy-resolved image of the real-space can be generated without being limited to a cross section in one dimension. This also allows an energy resolved diagonal section through the Brillouin zone, like shown in Figure 2c), without having to rotate the sample.

**Supporting Information**

Supporting Information is available from the Wiley Online Library or from the author.

**Acknowledgements**

S.B., J.D., M.D., M.E, S.H., and S.K. acknowledge financial support by the German Research Foundation (DFG) under Germany's Excellence Strategy–EXC2147 "ct.qmat" (project id 390858490) and HO 5194/12-1.

**Conflict of Interest**

The authors declare no conflict of interest.

**References**

[1] G. B. Jo, J. Guzman, C. K. Thomas, P. Hosur, A. Vishwanath, and D. M. Stamper-Kurn, Ultracold atoms in a tunable optical kagome lattice, *Phys. Rev. Lett.* **2012**, *108*, 045305.

[2] L. Tang, D. Song, S. Xia, S. Xia, J. Ma, W. Yan, Y. Hu, J. Xu, D. Leykam, and Z. Chen, Photonic flat-band lattices and unconventional light localization, *Nanophotonics* **2020**, *9*, 1161-1176.

[3] M. Polini, F. Guinea, M. Lewenstein, H. C. Manoharan, and V. Pellegrini, Artificial honeycomb lattices for electrons, atoms and photons, *Nature nanotechnology* **2013**, *8*, 625-633 (2013).

[4] I. Bloch, J. Dalibard, and W. Zwerger, Many-body physics with ultracold gases, *Rev. Mod. Phys.* **2008**, *80*, 885.

[5] F. Schäfer, T. Fukuhara, S. Sugawa, Y. Takasu, and Y. Takahashi, Tools for quantum simulation with ultracold atoms in optical lattices, *Nature Reviews Physics* **2020**, *2*, 411–425.

[6] R.C. Sterling, H. Rattanasonti, S. Weidt, K. Lake, P. Srinivasan, S.C. Webster, M. Kraft, and W.K. Hensinger, Fabrication and operation of a two-dimensional ion-trap lattice on a high-voltage microchip, *Nat. Commun.* **2014**, *5*, 3637.

[7] D. N. Neshev, T. J. Alexander, E. A. Ostrovskaya, Y. S. Kivshar, H. Martin, I. Makasyuk, and Z. Chen, Observation of Discrete Vortex Solitons in Optically Induced Photonic Lattices, *Phys. Rev. Lett.* **2004**, *92*, 123903.

[8] I. L. Garanovich, S. Longhi, A. A. Sukhorukov, and Y. S. Kivshar, Light propagation and localization in modulated photonic lattices and waveguides, *Physics Reports* **2012**, *518*, 1-79.

[9] L. Tang, D. Song, S. Xia, S. Xia, J. Ma, W. Yan, Y. Hu, J. Xu, D. Leykam and Z. Chen, Photonic flat-band lattices and unconventional light localization, *Nanophotonics* **2020**, *9*, 1161-1176.

[10] A. Amo and J. Bloch, Exciton-polaritons in lattices: A non-linear photonic simulator, *C. R. Physique* **2016**, *17*, 934.

[11] C. Schneider, K. Winkler, M. D. Fraser, M. Kamp, Y. Yamamoto, E. A. Ostrovskaya, and S. Höfling, Exciton-Polariton Trapping and Potential Landscape Engineering, *Rep. Prog. Phys.* **2017**, *80*, 016503.

[12] N. G. Berloff, M. Silva, K. Kalinin, A. Askitopoulos, J. D. Töpfer, P. Cilibrizzi, W. Langbein, and P. G. Lagoudakis, Realizing the classical XY Hamiltonian in polariton simulators, *Nat. Mater.* **2017**, *16*, 1120–1126.

[13] C. Weisbuch, M. Nishioka, A. Ishikawa, and Y. Arakawa, Observation of the coupled exciton-photon mode splitting in a semiconductor quantum microcavity. *Phys. Rev. Lett.* **1992**, *69*, 3314.

[14] H. Deng, H. Haug, and Y. Yamamoto, Exciton-polariton Bose-Einstein condensation, *Rev. Mod. Phys.* **2010**, *82*, 1489.

[15] J. Kasprzak, M. Richard, S. Kundermann, A. Baas, P. Jeambrun, J. M. J. Keeling, F. M. Marchetti, M. H. Szymańska, R. André, J. L. Staehli, V. Savona, P. B. Littlewood, B. Deveaud, and S. LeDang, Bose-Einstein condensation of exciton polaritons, *Nature* 443, 409–414 (2006).

[16] R. Balili, V. Hartwell, D. Snoke, L. Pfeiffer, and K. West, Bose–Einstein Condensation of Microcavity Polaritons in a Trap, *Science* **2007**, *316*, 1007– 1010.

[17] A. Amo, J. Lefrère, S. Pigeon, C. Adrados, C. Ciuti, I. Carusotto, R. Houdré, E. Giacobino, and A. Bramati, Superfluidity of polaritons in semiconductor microcavities, *Nat. Phys.* **2009**, *5*, 805-810.

[18] C. Schneider, A. Rahimi-Iman, N. Y. Kim, J. Fischer, I. G. Savenko, M. Amthor, M. Lermer, A. Wolf, L. Worschech, V. D. Kulakovskii, I. A. Shelykh, M. Kamp, S. Reitzenstein, A. Forchel, Y. Yamamoto, and S. Höfling, An electrically pumped polariton laser, Nature **2013**, *497*, 348–352.

[19] I. Carusotto and C. Ciuti, Quantum fluids of light, *Rev. Mod. Phys.* **2013**, *85*, 299.

[20] P. St-Jean, V. Goblot, E. Galopin, A. Lemaître, T. Ozawa, L. LeGratiet, I. Sagnes, J. Bloch, and A. Amo, Lasing in topological edge states of a one-dimensional lattice, *Nat. Photon.* **2017**, *11*, 651–656.

[21] T. H. Harder, M. Sun, O. A. Egorov, I. Vakulchyk, J. Beierlein, P. Gagel, M. Emmerling, C. Schneider, U. Peschel, I. G. Savenko, S. Klembt, and S. Höfling, Coherent Topological Polariton Laser, *ACS Photonics* **2021**, *8*, 1377–1384.

[22] K. Kusudo, N. Y. Kim, A. Löffler, S. Höfling, A. Forchel, Y. Yamamoto, Stochastic Formation of Polariton Condensates in Two Degenerate Orbital States, *Phys. Rev. B*, **2013**, 87, 214503.

[23] T. Jacqmin, I. Carusotto, I. Sagnes, M. Abbarchi, D. Solnyshkov, G. Malpuech, E. Galopin, A. Lemaître, J. Bloch, and A. Amo, Direct observation of Dirac cones and a flatband in a honeycomb lattice for polaritons, *Phys. Rev. Lett.* **2014**, *112*, 11.

[24] C. E. Whittaker, T. Dowling, A. V. Nalitov, A. V. Yulin, B. Royall, E. Clarke, M. S. Skolnick, I. A. Shelykh, and D. N. Krizhanovskii, Optical analogue of Dresselhaus spin–orbit interaction in photonic graphene, *Nat. Phys.* **2021**, *15*, 193-196.

[25]     H. Suchomel, S. Klembt, T. H. Harder, M. Klaas, O. A. Egorov, K. Winkler, M. Emmerling, R. Thomale, S. Höfling, and C. Schneider, Platform for Electrically Pumped Polariton Simulators and Topological Lasers, *Phys. Rev. Lett.* **2018**, *121*, 257402.

[26]     S. Klembt, T.H. Harder, O. A. Egorov, K. Winkler, R. Ge, M. A. Bandres, M. Emmerling, L. Worschech, T. C. H. Liew, M. Segev, C. Schneider, and S. Höfling, Exciton-polariton topological insulator, *Nature* **2018**, *562*, 552–556.

[27]     S. I. Tsintzos, N. T. Pelekanos, G. Konstantinidis, Z. Hatzopoulos, and P. G. Savvidis, A GaAs polariton light-emitting diode operating near room temperature, *Nature* **2008**, *453*, 372-375.

[28]     P. Gagel, T. H. Harder, S. Betzold, O. A. Egorov, J. Beierlein, H. Suchomel, M. Emmerling, A. Wolf, U. Peschel, S. Höfling, C. Schneider, and S. Klembt, Electro-optical Switching of a Topological Polariton Laser, *ACS Photonics* **2022**, *9*, 405–412.

[29]     S. Kéna-Cohen and S. R. Forrest, Room-temperature polariton lasing in an organic single-crystal microcavity, *Nat. Photonics* **2010**, *4*, 371.

[30]     J. D. Plumhof, T. Stöferle, L. Mai, U. Scherf, and R. F. Mahrt, Room- temperature Bose–Einstein condensation of cavity exciton–polaritons in a polymer, *Nat. Mater.* **2014**, *13*, 247 (2014).

[31]     C. P. Dietrich, A. Steude, L. Tropf, M. Schubert, N. M. Kronenberg, K. Ostermann, S. Höfling, M. C. Gather, An exciton-polariton laser based on biologically produced fluorescent protein, *Sci. Adv.*, **2016**, *2*, e1600666.

[32]     J. Keeling and S. Kéna-Cohen, Bose–Einstein Condensation of Exciton- Polaritons in Organic Microcavities, *Annu. Rev. Phys. Chem.* **2020**, *71*, 435.

[33]     X. Liu, T. Galfsky, Z. Sun, F. Xia, E.-C. Lin, Y.-H. Lee, S. Kéna-Cohen, and V. M. Menon, Strong light–matter coupling in two-dimensional atomic crystals. Nat. Photon. **2015,** *9*, 30–34.

[34]     S. Dufferwiel, S. Schwarz, F. Withers, A. A. P. Trichet, F. Li, M. Sich, O. Del Pozo-Zamudio, C. Clark, A. Nalitov, D. D. Solnyshkov, G. Malpuech, K. S. Novoselov, J. M. Smith, M. S. Skolnick, D. N. Krizhanovskii, and A. I. Tartakovskii, Exciton–polaritons in van der Waals heterostructures embedded in tunable microcavities. Nat. Commun. **2015**, *6*, 8579.

[35]     N. Lundt, S. Klembt, E. Cherotchenko, S. Betzold, O. Iff, A. V. Nalitov, M. Klaas, C. P. Dietrich, A. V. Kavokin, S. Höfling, and C. Schneider, Room-temperature Tamm-plasmon exciton–polaritons with a WSe2 monolayer. Nat. Commun. **2016**, *7*, 13328.

[36]     K. F. Mak and J. Shan, Photonics and optoelectronics of 2D semiconductor transition metal dichalcogenides. Nat. Photon. **2016**, *10*, 216–226.

[37]     A. Brehier, R. Parashkov, J.-S. Lauret, and E. Deleporte, Strong exciton-photon coupling in a microcavity containing layered perovskite semiconductors. *Appl. Phys. Lett.* **2006**, *89*, 171110.

[38]     R. Su, C. Diederichs, J. Wang, T. C. H. Liew, J. Zhao, S. Liu, W. Xu, Z. Chen, and Q. Xiong, Room-temperature polariton lasing in all-inorganic perovskite nanoplatelets. *Nano Lett.* **2017**, *17*, 3982–3988.

[39]     A. Fieramosca, L. Polimeno, V. Ardizzone, L. De Marco, M. Pugliese, V. Maiorano, M. De Giorgi, L. Dominici, G. Gigli, D. Gerace, D. Ballarini, and D. Sanvitto, Two-dimensional hybrid perovskites sustaining strong polariton interactions at room temperature. *Sci. Adv.* **2019**, *5*, eaav9967.

[40]     K. Łempicka-Mirek, M. Król, H. Sigurdsson, A. Wincukiewicz, P. Morawiak, R. Mazur, M. Muszyński, W. Piecek, P. Kula, T. Stefaniuk, M. Kamińska, L. De Marco, P. G. Lagoudakis, D. Ballarini, D. Sanvitto, J. Szczytko, B. Piętka, Electrically tunable Berry curvature and strong light-matter coupling in liquid crystal microcavities with 2D perovskite, *Sci. Adv.* **2022**, *8*, eabq7533.

[41]     R. Jayaprakash, C. E. Whittaker, K. Georgiou, O. S. Game, K. E. McGhee, D. M. Coles, and D. G. Lidzey, Two-Dimensional Organic-Exciton Polariton Lattice Fabricated Using Laser Patterning, *ACS Photonics* **2020**, *7*, 2273–2281.

[42]     F. Scafirimuto, D. Urbonas, M. A. Becker, U. Scherf, R. F. Mahrt, and T. Stöferle, Tunable exciton–polariton condensation in a two-dimensional Lieb lattice at room temperature, *Communications Physics* **2021**, *4*, 39.

[43]     J. Wu, S. Ghosh, Y. Gan, Y. Shi, S. Mandal, H. Sun, B. Zhang, T. C. H. Liew, R. Su, Q. Xiong, Higher-order topological polariton corner state lasing. Sci. Adv. **2023**, *9*, eadg4322.

[44]     O. Shimomura, The discovery of aequorin and green fluorescent protein. Journal of Microscopy **2005**, *217*, 3-15.

[45]     M. Dusel, S. Betzold, O. A. Egorov, Room temperature organic exciton–polariton condensate in a lattice, *Nat. Commun.* **2020**, *11*, 2863.

[46]     M. Dusel, S. Betzold, T. H. Harder, M. Emmerling, J. Beierlein, J. Ohmer, U. Fischer, R. Thomale, C. Schneider, S. Höfling, and S. Klembt, Room-Temperature Topological Polariton Laser in an Organic Lattice, Nano Lett. **2021**, *21*, 6398–6405.

[47]     A. H. Castro Neto, F. Guinea, N. M. R. Peres, K. S. Novoselov, and A. K. Geim, The electronic properties of graphene, *Rev. Mod. Phys.* **2009**, *81*, 109.

[48] S. Y. Zhou, G.-H. Gweon, A. V. Fedorov, P. N. First, W. A. de Heer, D.-H. Lee, F. Guinea, A. H. Castro Neto & A. Lanzara, Substrate-induced bandgap opening in epitaxial graphene, *Nature Materials* **2007**, *6*, 770–775.

[49] S. Klembt, T. H. Harder, O. A. Egorov, K. Winkler, H. Suchomel, J. Beierlein, M. Emmerling, C. Schneider, S. Höfling, Polariton condensation in S- and P-flatbands in a two-dimensional Lieb lattice, *Appl. Phys. Lett.* **2017**, *111*, 231102.

[50] C. E. Whittaker, E. Cancellieri, P. M. Walker, D. R. Gulevich, H. Schomerus, D. Vaitiekus, B. Royall, D. M. Whittaker, E. Clarke, I. V. Iorsh, I. A. Shelykh, M. S. Skolnick, and D. N. Krizhanovskii, Exciton Polaritons in a Two-Dimensional Lieb Lattice with Spin-Orbit Coupling, *Phys. Rev. Lett.* **2018**, *120*, 097401.

[51] T. H. Harder, O. A. Egorov, C. Krause, J. Beierlein, P. Gagel, M. Emmerling, C. Schneider, U. Peschel, S. Höfling, and S. Klembt, Kagome Flatbands for Coherent Exciton-Polariton Lasing, *ACS Photonics* **2021**, *8*, 3193–3200.

[52] B. Real, O. Jamadi, M. Milićević, N. Pernet, P. St-Jean, T. Ozawa, G. Montambaux, I. Sagnes, A. Lemaître, L. Le Gratiet, A. Harouri, S. Ravets, J. Bloch, and A. Amo, Semi-Dirac Transport and Anisotropic Localization in Polariton Honeycomb Lattices, *Phys. Rev. Lett.* **2020**, *125*, 186601.

[53] T. H. Harder, PhD Thesis, University of Würzburg, urn:nbn:de:bvb:20-opus-259008, February, **2022**.

[54] S. Betzold, M. Dusel, O. Kyriienko, C. P. Dietrich, S. Klembt, J. Ohmer, U. Fischer, I. A. Shelykh, C. Schneider, and S. Höfling, Coherence and Interaction in Confined Room-Temperature Polariton Condensates with Frenkel Excitons, *ACS Photonics* **2020**, *7*, 384– 392.

[55] T. Yagafarov, D. Sannikov, A. Zasedatelev. K. Georgiou, A. Baranikov, O. Kyriienko, I. Shelykh, L. Gai, Z. Shen, D. Lidzey, and P. Lagoudakis Mechanisms of blueshifts in organic polariton condensates, *Commun. Phys.* **2020**, *3*, 18.

[56] N. Y. Kim, K. Kusudo, C. Wu, N. Masumoto, A. Löffler, S. Höfling, N. Kumada, L. Worschech, A. Forchel, and Y. Yamamoto, Dynamical d-wave condensation of exciton–polaritons in a two-dimensional square-lattice potential, *Nat. Phys.* **2011**, *7*, 681.

[57] L. Lackner, M. Dusel, O. A. Egorov, B. Han, H. Knopf, F. Eilenberger, S. Schröder, K. Watanabe, T. Taniguchi, S. Tongay, C. Anton-Solanas, S. Höfling, and C. Schneider, Tunable exciton-polaritons emerging from WS2 monolayer excitons in a photonic lattice at room temperature. *Nat. Commun.* **2021**, *12*, 4933.

[58]　N. Tomm, A. Javadi, N. O. Antoniadis, D. Najer, M. C. Löbl, A. R. Korsch, R. Schott, S. R. Valentin, A. D. Wieck, A. Ludwig, and R. J. Warburton, A bright and fast source of coherent single photons, *Nat. Nanotechnol.* **2021**, *16*, 399-403.

**ToC**

This study presents a precisely controlled two-dimensional optical lattice for the study of exciton-polaritons at room-temperature, using a fluorescent protein in a microcavity to achieve excellent cavity quality and showcase polariton lasing. Providing insights into nonlinear many-body physics and bosonic condensation, with potential applications e.g. in topological lasers, this work marks a notable step forward in the advancement of room temperature photonic and polaritonic lattice physics, opening avenues for diverse optoelectronic devices.

Simon Betzold*, Johannes Düreth, Marco Dusel, Monika Emmerling, Antonina Bieganowska, Jürgen Ohmer, Utz Fischer, Sven Höfling, and Sebastian Klembt*

**Dirac Cones and Room Temperature Polariton Lasing Evidenced in an Organic Honeycomb Lattice**

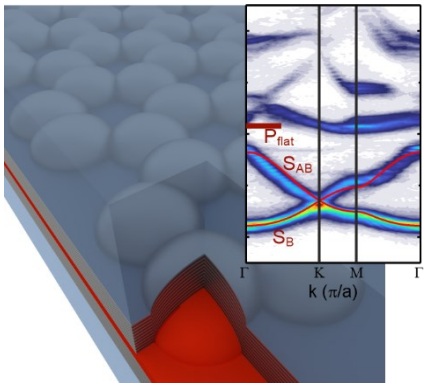

# Supporting Information

**Dirac Cones and Room Temperature Polariton Lasing Evidenced in an Organic Honeycomb Lattice**

*Simon Betzold\*, Johannes Düreth, Marco Dusel, Monika Emmerling, Antonina Bieganowska, Jürgen Ohmer, Utz Fischer, Sven Höfling, and Sebastian Klembt\**

S. Betzold, J. Düreth, M. Dusel, M. Emmerling, S. Höfling, S. Klembt

Julius-Maximilians-Universität Würzburg, Physikalisches Institut and Würzburg-Dresden Cluster of Excellence ct.qmat, Lehrstuhl für Technische Physik, Am Hubland, 97074 Würzburg, Germany

A. Bieganowska

Wroclaw University of Science and Technology, Faculty of Fundamental Problems of Technology, Department of Experimental Physics, Wyb. Wyspiańskiego 27, 50-370 Wroclaw, Poland

J. Ohmer, U. Fischer

Julius-Maximilians-Universität Würzburg, Department of Biochemistry, Am Hubland, 97074 Würzburg, Germany

\*E-mail: simon.betzold@uni-wuerzburg.de, sebastian.klembt@uni-wuerzburg.de

## S1: AFM-measurements of the structured DBR

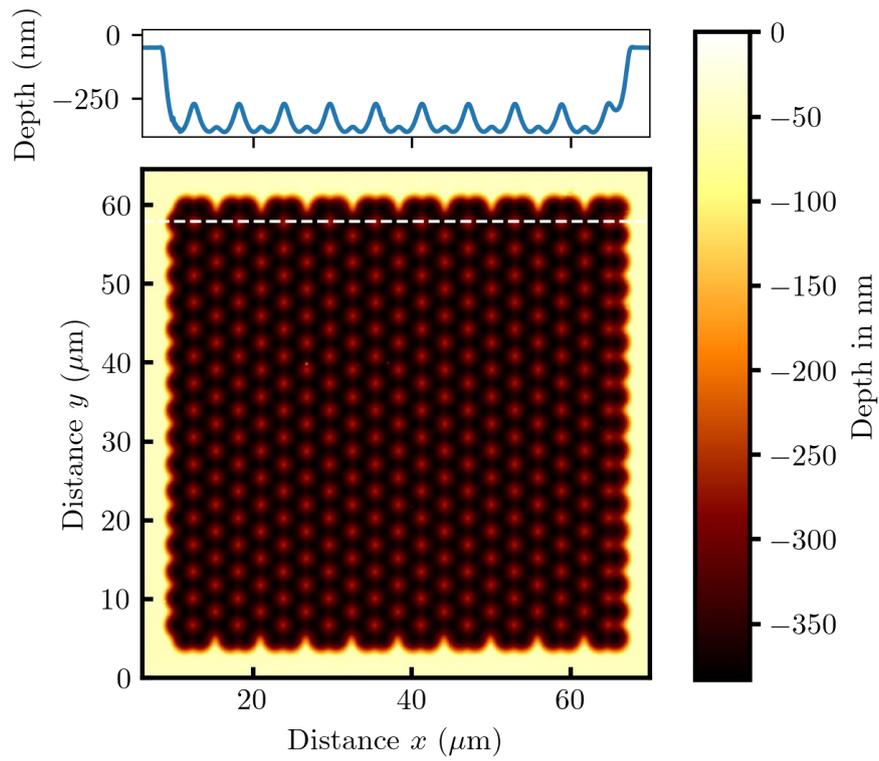

**Figure S1.** Visualization of the photonic potential landscape illustrated by an AFM image of the honeycomb lattice. The sites have a diameter of 4 µm, a depth of about 380 nm and a center-to-center distance of 2 µm.

## S2: Sample assembly

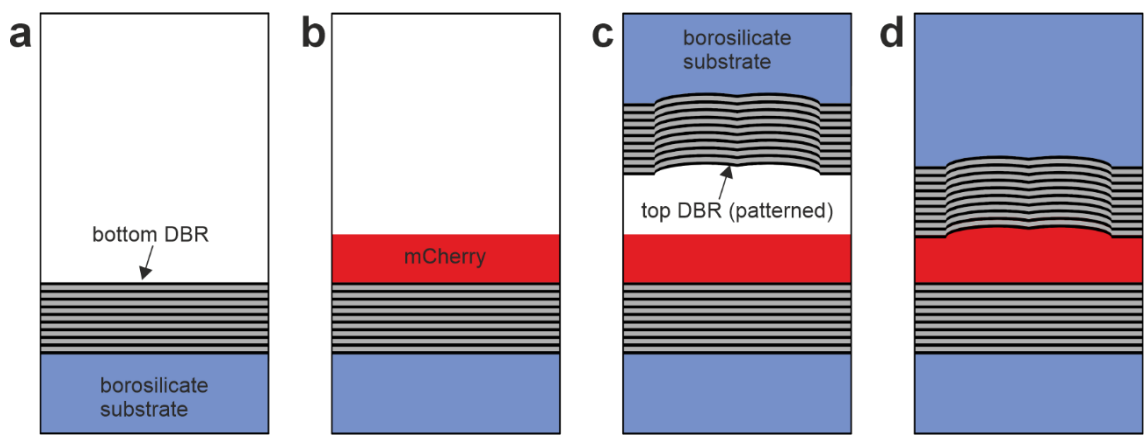

**Figure S2.** Schematic representation of the sample assembly. Starting with a planar DBR consisting of alternating SiO2/TiO2 layers (a), a solution of mCherry in water is pipetted on top and spread (b). Finally, the second, patterned DBR (c) is applied to the mCherry (d), pressed and left to dry under constant pressure for 48 hours.

## S3: Determination of the Rabi splitting, the cavity length, and the Q-factor

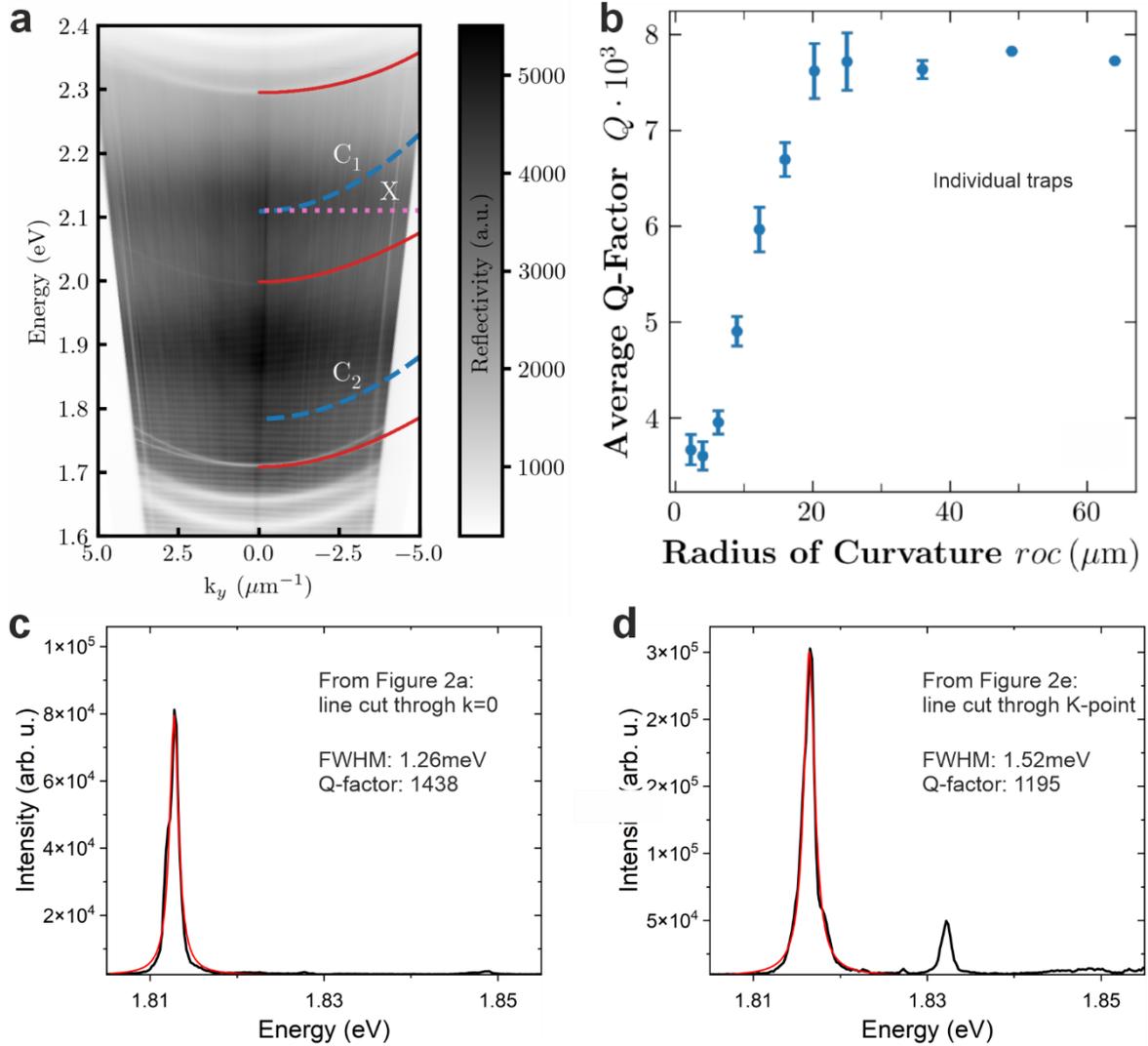

**Figure S3:** Reflection measurement (a) of the planar cavity near the measured honeycomb lattice of the main text. C1 and C2 denote the photonic modes, while X denotes the excitonic mode. The fitted polaritonic resonances are shown in red. In b) the measured average Q-factor is plotted for potentials with different radii of curvature. Individual hemispheres of the measured honeycomb lattice have a radius of curvature of 10.5 um. c) and d) show spectra taken from line cuts of Figure 2a through k=0 and Figure 2e at the K-point, respectively. The Q-factor of the binding s-band is about 1400, while the Q-factor at the K-point is about 1200.

To determine the cavity length and Rabi splitting of the sample, three independent white light reflection measurements at different detunings were evaluated. The spectral position of the photonic modes C1 and C2 was determined from transfer matrix simulations of the empty cavity and iteratively updated to find the Rabi splitting at the three positions. Our results yield an average Rabi splitting of (313.8±4.7) meV and optical thicknesses for the mCherry layer of about 1100 nm, 1150 nm, and 1080 nm. As expected, the Rabi splitting is almost constant over this detuning range, so we show the measurement in Figure S3a at an optical thickness of mCherry of about 1100 nm and a Rabi splitting of 318 meV.

We include Figure S3b to illustrate the dependence of the Q-factor on the radius of curvature of the structures. The data shown in this figure are measured on a different sample but using similar mirrors and structures. The Q-factor reaches a maximum of about 8000 for radii of curvature larger than 30 µm but decreases rapidly below this value as scattering losses increase and the stability condition for stable resonators is approached.

In addition, Figure S3c and d show that the Q-factor decreases further in a lattice of coupled resonators due to increased scattering losses.

## S4: Mode tomography in the linear regime of the mode analyzed in Figure 3

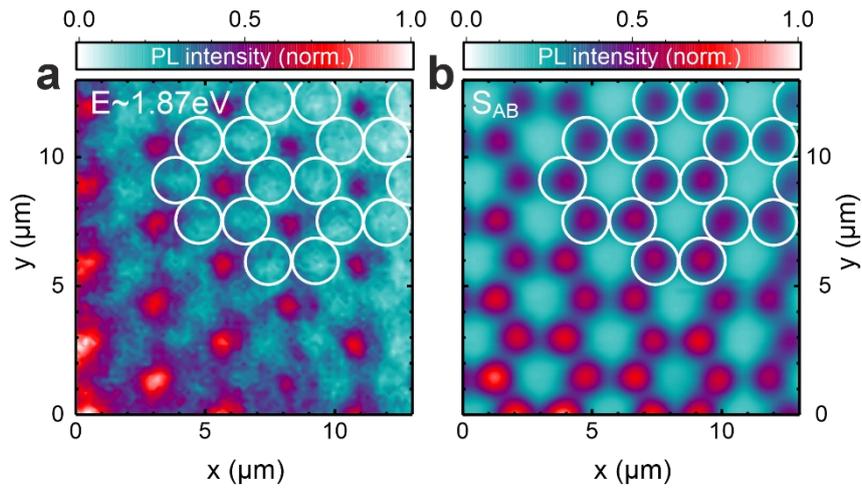

**Figure S4:** (a) Real-space image, for which the intensity was spectrally summed between 1.869 eV and 1.872 eV. The origin of these specific Bloch mode is most likely due to the interference of the $3d_{xy}$ modes of the individual hemisphere potentials. For this reason, the center of each hexagon shows the highest emission intensity, while the part of the modes located in the center of each hemisphere interferes destructively. A similar signature was observed by Kim et al. in a square lattice in an inorganic microcavity.[56] (b) For comparison: Real-space image of the anti-binding S subband, for which the intensity was spectrally summed between 1.818 eV and 1.830 eV (same as Figure 2h) of the main text).

## S5: Additional information about the experimental setup

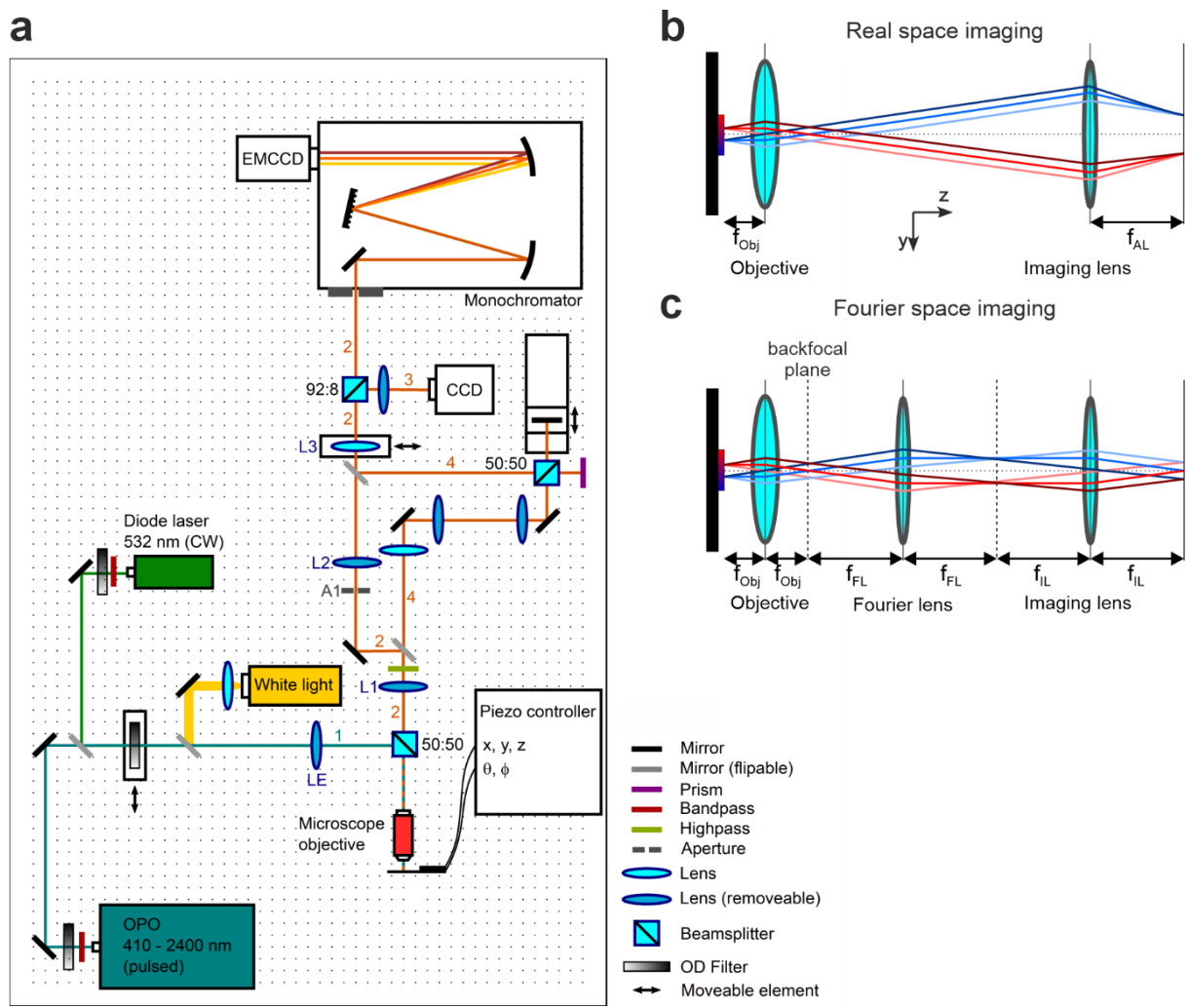

**Figure S5.** Schematic drawing of the measurement setup with the most important components for photoluminescence and reflection measurements (a) and schematic lens configuration to measure the real space (b) and the Fourier space (c).

Various laser systems are available for measurements, directed through (flip) mirrors onto the excitation path (Figure S5, Path 1). For basic sample characterization, a frequency-doubled ND:YAG laser diode with an emission wavelength of 532 nm (Atelier Rieter, DPSS module, 200 mW output power, CW) is used. Additionally, an optical parametric oscillator (OPO) from OPOTEC Inc. is available for measurements in the nonlinear regime, tunable from 410 to 2400 nm, with pulse rates up to 20 Hz, a pulse length of 7 ns, and wavelength-dependent output power up to 9 mJ. The excitation power to the sample can be varied using a movable continuous OD filter. A white-light source can also be coupled into the excitation path using a flip mirror.

The emission from the excitation sources is directed from the excitation path via a 50:50 beam splitter to a microscope objective with 50x magnification and a numerical aperture of NA=0.42, focusing it onto the sample. A lens (Lens LE) can be introduced into the beam path, allowing measurements with an expanded laser spot. The emitted or reflected light is collected by the same microscope objective, passes through the beam splitter again, and is directed to the detection path (Path 2). A lens system leads to a Czerny-Turner spectrometer (Andor Solis Shamrock 500i), which spectrally filters the signal and directs it to a Peltier-cooled CCD camera. The CCD camera (Andor Newton 971) has an EM sensor and a resolution of 1600 x 400 pixels with a pixel size of 16 μm x 16 μm. This configuration allows a maximum nominal spectral resolution of about 200 μeV in the relevant energy range around 2 eV. Additionally, a second CCD camera without spectral resolution capability is accessible through Path 3.

The lens system in detection Path 2 enables measurements in both real-space and momentum-space (Fourier space). The two measurement methods are schematically depicted in Figure S5b and c. The microscope objective is simplified here as a single lens. For real-space, only Lens L3 from Figure S5a is needed, functioning as an imaging lens and focusing the real-space image onto the entrance slit of the monochromator. The magnification of the real-space image is given by the ratio of the focal length of the imaging lens $f_{IL}$ to the focal length of the objective $f_{Obj}$. The chip of the CCD camera displays the image of the sample with an open entrance slit and at a grating position in $0^{th}$ reflection order. Closing the entrance slit and rotating the grating allows energy resolution, reducing the spatial information to the x-direction while $y \approx 0$.

For measurements in momentum-space, the Fourier lens L1 is introduced. Taking advantage of the fact that all rays emitted from the sample at a certain angle converge to a point in the back-focal plane of the objective, Lens L1 and Lens L3 project the Fourier plane onto the entrance slit of the monochromator. Closing the entrance slit to about 30 μm reduces the analyzed light to wave vectors around $k_y \approx 0$. An intermediate real space image is formed between the Fourier lens and the imaging lens, and the aperture A1 in this plane allows selection of the spatial region for the k-resolved measurement.

Lens L3 is mounted on a motorized linear stage, enabling consecutive sections of the real or momentum space at a fixed y or $k_y$ value to be captured and then assembled. This allows, for example, the generation of an energy-resolved image of the real space without limiting it to a cut in one dimension. This measurement method is referred to as mode tomography in the main text. To provide a conceptual distinction from measurements in momentum space, the term hyperspectral imaging is used here.

A flip mirror in detection path 2 can guide the emitted radiation back into the detection path via a Michelson interferometer (Path 4). This allows measurements of the spatial (and temporal) first-order correlation function.

**References**

[56] N. Y. Kim, K. Kusudo, C. Wu, N. Masumoto, A. Löffler, S. Höfling, N. Kumada, L. Worschech, A. Forchel, and Y. Yamamoto, Dynamical d-wave condensation of exciton–polaritons in a two-dimensional square-lattice potential, *Nat. Phys.* **2011**, *7*, 681.